\documentstyle[11pt,adassconf]{article}  

\begin{document}   

\paperID{08-01}

\title{The SDSS Imaging Pipelines}

\author{Robert Lupton, James E. Gunn, \v{Z}eljko Ivezi\'c, Gillian R. Knapp}
\affil{Princeton University Observatory
Princeton University,
Princeton, NJ~08544
}
\author{Stephen Kent}
\affil{Fermi National Accelerator Laboratory,
Batavia,
Il}
\author{Naoki Yasuda}
\affil{National Astronomical Observatory, 2-21-1, Mitaka,
	Tokyo, 181-0015, Japan}

\contact{Robert Lupton}
\email{rhl@astro.princeton.edu}

\paindex{Lupton, R. H.}
\aindex{Gunn, J. E.}
\aindex{Ivezi\'c, Z.}
\aindex{Knapp, G. R.}
\aindex{Kent, S.}
\aindex{Yasuda, N.}



\begin{abstract}          

We summarise the properties of the Sloan Digital Sky Survey (SDSS)
project, discuss our software infrastructure, and outline the
architecture of the SDSS image processing pipelines.

We then discuss two of the algorithms used in the SDSS image processing;
the Karhunen-Lo\`eve transform based modelling of the spatial variation of the PSF,
and the use of galaxy models in star/galaxy separation.

We conclude with the first author's personal opinions on the challenges
that the astronomical community faces with major software projects.

\end{abstract}


\section{Introduction}

The SDSS (York et al. 2000) consists of four major components: a
dedicated 2.5m telescope at Apache Point, New Mexico, along with a
separate 50cm telescope used to monitor the extinction and to provide
calibration patches for the main telescope; a large format imaging
camera (Gunn et al. 1998) containing 30 $2048\times2048$ (13$\times$13
arcmin) photometric CCDs with $u' g' r' i' z'$ filters
and 24 $2048\times400$ astrometric and focus CCDs; two 320-fibre-fed double
spectrographs, each with two $2048\times2048$ CCDs; and lots and lots
of software, with contributions from most of the SDSS institutions (listed
in the acknowledgments).

The primary goals of the project are to survey the Northern Galactic
Cap ($\approx 10^4$ square degrees) in five bands to
(PSF) limits of 22.3(u'), 23.3(g'), 23.1(r'), 22.3(i'), and 20.8(z'),
and to carry out
a spectroscopic survey of $10^6$ galaxies, $10^5$ QSOs, and a few
$\times 10^4$ stars.

The SDSS is now in operational mode, and as of this writing (late
January 2001) has imaged some 1600 $\rm deg^2$ and obtained about
120,000 spectra as part of its commissioning and initial operations
phases.  These data have allowed dramatic new astronomical discoveries
to be made, discoveries that we shall not further discuss here
(e.g. Blanton et al. 2001, Fan et al. 2000, 2001, Fischer et al. 2000,
Ivezi\'c et al. 2000, Leggett et al. 2000).

\section{Software Infrastructure}

\subsection{Configuration Management and Bug Reporting}

The SDSS took an early decision to use public domain software wherever
possible; in practice this has largely been applied to our infrastructure
rather than scientific codes.

Our software engineering tools are entirely public domain (with the exception
of compilers).

We adopted \htmladdnormallinkfoot{cvs}{http://www.cvshome.org/}
as a source code manager and have been pleased
with its performance.  We currently have about 1.7Gby in our cvs
repository (including at least one version of IRAF).  We have found
that, after an initial period of distrust, scientists have found cvs
to be extremely useful; in at least some cases, people sitting next to
each other at the observatory in New Mexico have communicated via a cvs repository in
Illinois.

While cvs allows us to control individual pieces of software, it does not
provide a means of controlling complete systems.  We have used a
Fermi National Accelerator Laboratory (FNAL) utility called
\htmladdnormallinkfoot{ups}{http://www.fnal.gov/docs/products/ups/} which
allows us to associate a set of \textit{dependent products} with a
piece of our software.  For example, version \texttt{v5\_2\_10} of the
image processing pipeline depends upon  \texttt{v7\_15} of our infrastructure
routines.  This enables us to guarantee that at any time in the future
we can reconstruct an entire system, using exactly the same bits and pieces.
The particular versions (e.g. \texttt{v5\_2\_10}) correspond to tags in the
cvs repository. We have adopted a procedure that stable versions of
our pipelines correspond to \textit{branch tags} in cvs; this has allowed
us to proceed with development while giving us the ability to fix bugs
found in the stable, delivered, code.

We have used
\htmladdnormallinkfoot{gnats}{http://sources.redhat.com/gnats/} as our
problem report and bug database. Since July 1998 we have acquired 1799
entries in the database; the last thousand have been filed since
February 2000.

\begin{figure}
\epsscale{0.6}
\plotone{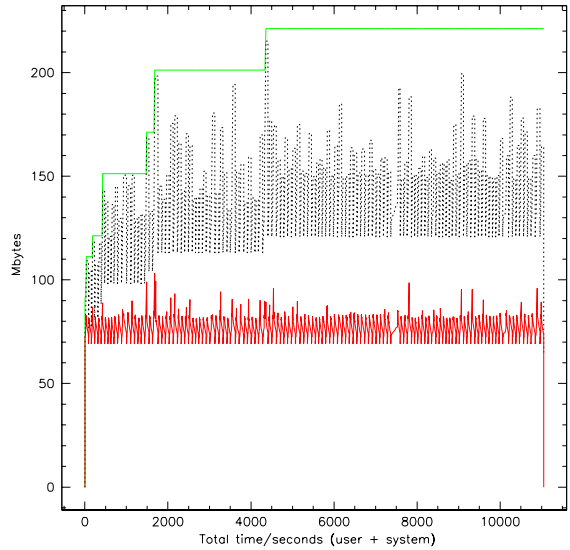}
\caption{A trace of the memory used while processing 121 fields (3.6Gb)
of an SDSS imaging run on a single 800MHz alpha processor.  A total of
165029 objects were detected and charcterised in 5 bands, giving a
rate of 13.4ms/object/band for processing from raw CCD frame to reduced
catalog.
\hfil\break
The figure has three lines illustrating memory usage versus time.
The lower line is the
memory actively in use; the middle dotted line shows the memory in
heap, and the top line shows the memory allocated from the system. 
The difference between the upper two lines is \textit{guaranteed} to be
in 10Mb blocks, all except one of which is completely unused, and
can safely be assumed to be swapped out to disk.
}
\label{figMemTrace}
\end{figure}

\subsection{Command Interpreter}

We use a heavily enhanced version of TCL 7.4 (actually, of TCLX) as our
command language.  Much of the work developing this system (known as
\textit{dervish}, n\'e \textit{shiva}, Sergey et al. 1996) was carried
out at FNAL.

In addition to what now appear to be basically cosmetic changes (which
we regret), the major enhancements that we made were:
\begin{itemize}
  \item Memory tracing/defragmentation/debugging

A common problem with programs that make heavy use of dynamically
allocated memory is that the memory acquired from the operating system
becomes \textit{fragmented}, or that the program forgets to free
resources.  Both of these problems can be resolved by adding a layer
above \texttt{malloc}, and we have done so.  Figure \ref{figMemTrace}
shows that the total memory used in the steady state by the frames
pipeline (see below) is well controlled.

\item Support for C datatypes at tcl level.

We wrote a processor that scans the C include files (`.h files') and
generated a description of the schema of all the types declared therein.
This was originally used to implement a primitive persistent store, but
proved more useful in making the C data elements available at the TCL
prompt; this greatly increased the power and flexibility of our command
language, allowing us to build the command-and-control parts of our
pipelines in TCL rather than having to use compiled C. For example,
\hfill\break{\small\tt
assert \{[exprGet \$c.calib<\$i>->filt<0>] == \$f\}\hfil\break
handleSet \$fieldparams.frame<\$i>.fullWell<0> \$fullWell(0,\$f)\hfil\break}
where a `handle' is an address and a datatype.

\item Easy(ish) bindings from C to tcl.

We implemented a set of library calls that made it possible to bind C
commands to TCL in a way that, if not simple, at least required no
thought and could be handled by pasting appropriate boiler-plate code.

\end{itemize}

If we were starting this problem today, we would probably not use TCL
(maybe python in its PyRAF \paperref{D-05} incarnation?), and we would
certainly make greater efforts to use \emph{vanilla}, up-to-date,
versions of our chosen system.

\section{Imaging Pipelines}

The SDSS has quite a large number of pipelines which must be run in order
to fully process the data; we shall not discuss the spectroscopic
reductions or the operational and scientific databases.

\begin{figure}
\epsscale{0.8}
\plotone{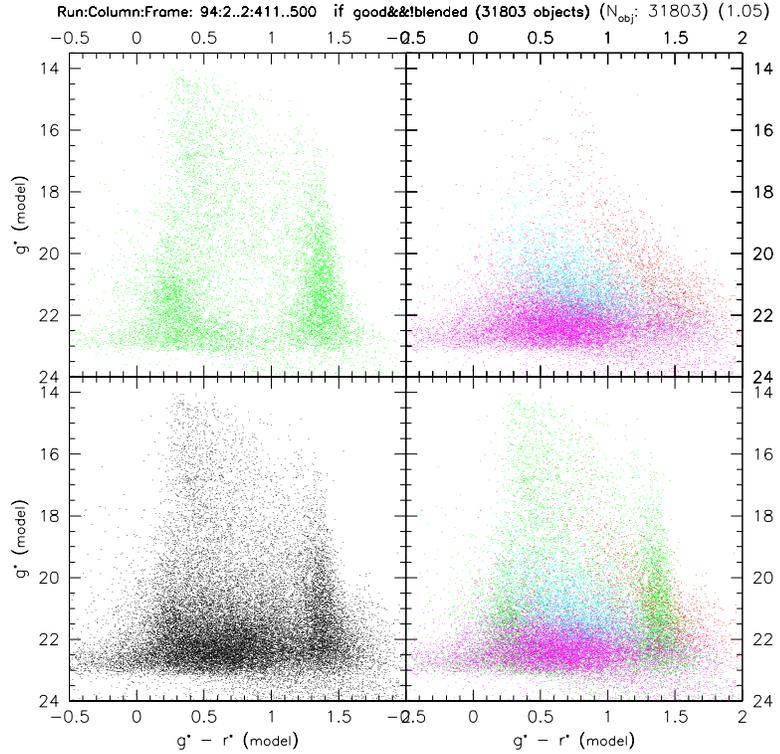}
\caption{A $g'$ v. $g'-r'$ colour-magnitude diagram containing 31803 objects
from SDSS commissioning data. The bottom two panels show all objects, the
top left shows only stars and the top right only galaxies. The disk and
halo turnoffs are clearly seen in the stellar diagram.
\hfil\break
If you are
viewing this figure in colour, green points
are stars; red points are galaxies classified morphologically as having
deVaucouleur-like profiles; cyan points have exponential profiles; and
magenta points are unclassified galaxies.  In black and white, the
bottom two panels are unfortunately indistinguishable}
\label{figGR}
\end{figure}

\begin{itemize}
\item \texttt{Astroline}

On the MVE167 processors (running vxWorks) used to archive the raw
data on the mountain, we also run a pipeline that 
processes the pixels before they're written to disk/tape. We generate
star cutouts (`Postage Stamps') and column quartiles; this is all that
we save from the 22 astrometric CCDs.

\item \texttt{MT} Pipeline.

Process the Photometric Telescope camera data. This consists of a set
of staring-mode observations of fields of standard stars, used to
define the extinction and photometric zero-points for the 2.5m scans.

\item \texttt{Serial Stamp Collecting} (SSC) Pipeline.

Reorganise the data stream, cut a more complete set of Postage Stamps.

\item \texttt{Astrometric} Pipeline.

Process the centroids of stars from \texttt{astroline}/\texttt{SSC}
and generate the astrometric transformations from pixels to
$(\alpha,\delta)_{J2000}$ and between bands.

\item \texttt{Postage Stamp} Pipeline (PSP).

Estimate the flat field vectors, bias drift, and sky levels,
and characterise the PSF for each field.

\item \texttt{Frames} Pipeline.

Process the full imaging data, producing corrected frames, object
catalogues, and atlas images.

\item Calibration.

Take the outputs from \texttt{MT} pipeline and \texttt{frames},
and convert counts to fluxes.

\end{itemize}

One major gain from splitting responsibilities in this way is that
once we get to the \texttt{frames} pipeline, fields (10arcmin$\times$14arcmin
patches on the sky) may be processed independently and in any order.

\section{Interesting Algorithms}

The SDSS imaging piplelines employ a number of novel and even
interesting algorithms, which are slowly being written up for
publication; for example the image deblender (Lupton 2001). Here we
shall only discuss a couple of features connected to handling the
point spread function (PSF) and the related problem of star/galaxy
separation.

\subsection{PSF Estimation}

Even in the absence of atmospheric inhomogeneities the SDSS telescope
delivers images whose FWHMs vary by up to 15\% from one side of a CCD
to the other; the worst effects are seen in the chips furthest from
the optical axis.

If the seeing were constant in time one might hope to understand these
effects ab initio, but when coupled with
time-variable seeing the delivered image quality is a complex
two-dimensional function and we chose to model it heuristically
using a Karhunen-Lo\`eve (KL) transform.

\subsubsection{Why We Need to Know the PSF}

The description of the PSF (as derived in the next subsection) is
critical for accurate PSF photometry, i.e. for all faint object
photometry --- if the PSF varies so does the aperture correction.

We also need to accurately know the PSF in order to be able to
separate stars from galaxies; after all, the \emph{only} valid
discriminant that isn't based on colours or priors is that galaxies
don't look like stars.

A good knowledge of the local PSF is also needed for all studies
that measure the shapes of non-stellar objects (e.g. weak lensing studies,
Fischer et al. 2000).

\subsubsection{KL Expansion of the PSF}

The first step is to identify a set of reasonably bright,
reasonably isolated stars from our image. 
We then use
these stars to form a KL basis, retaining the first $n$
terms of the expansion:
\begin{eqnarray}
P_{(i)}(u,v) = \sum_{r=1}^{r=n} a^r_{(i)} B_r(u,v)
\end{eqnarray}
where $P_{(i)}$ is the $i^{th}$ PSF star, the $B_r$ are the KL basis
functions, and $u,v$ are pixel coordinates relative to the origin of
the basis functions. In determining the $B_r$, the $P_{(i)}$ are
normalised to have equal peak value.

Once we know the $B_r$ we can write 
\begin{eqnarray}
a^r_{(i)} \approx \sum_{l = m = 0}^{l + m \le N} b^r_{lm} x_{(i)}^l y_{(i)}^m
\end{eqnarray}
where $x,y$ are the coordinates of the centre of the $i^{th}$ star, $N$
determines the highest power of $x$ or $y$ to include in the expansion,
and the $b^r_{lm}$ are determined by minimising
\begin{eqnarray}
\sum_i \left(P_{(i)}(u,v) - \sum_{r=1}^{r=n} a^r_{(i)} B_r(u,v)\right)^2;
\end{eqnarray}
note that all stars are given equal weight as we are interested in
determining the spatial variation of the PSF, and do not want to tailor
our fit to the chance positions of bright stars.

\subsubsection{Application to SDSS data}

For each CCD, in each band, there are typically 15-25
stars in a frame that we can use to determine the PSF; we usually
take $n = 3$ and $N = 2$ (i.e. the PSF spatial variation is
quadratic).  We need to estimate $n$ KL basis images, and a total of
$n(N + 1)(N+2)/2$ $b$ coefficients, and at first sight the problem might seem
underconstrained.  Fortunately we have many \textit{pixels} in each
of the $P_{(i)}$, and thus only the number of spatial terms ($(N + 1)(N+2)/2$,
i.e. 6 for $N = 2$) need be compared with the number of stars available.

In fact, rather than use only the stars from a single frame to determine
that frame's PSF, we include stars from both proceeding and succeeding frames
in the fit.  This procedure has several advantages: the spatial variation is better
constrained at the leading and trailing edges of the frame; the
PSF variation is smoother from frame to frame; and the PSF is determined
from more stars.

We have found that optimal results are obtained by using a range of
$\pm 2$ frames to determine the KL basis functions
$B_r$ and $\pm 1/2$ frame to follow the spatial variation of the
PSF. If we try to use a larger window we find that variation of the $a^r$
coefficients is not well described by the polynomials that we have
assumed.  We have not tried using a different set of expansion functions
(e.g. a Fourier series).

\subsection{Model Fitting and Star/Galaxy Separation}

\begin{figure}
\epsscale{0.5}
\plotone{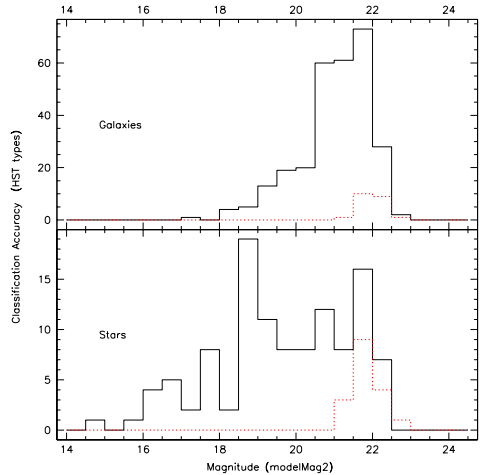}
\caption{
Star-Galaxy separation in the SDSS. The bottom panel shows object that
are classified as stars based on their HST morphology; the top panel
shows galaxies. The x-axis is the $r'$ model magnitude.
The solid line shows the number of objects classified
correctly by the SDSS pipeline, the (red) dotted line shows the objects
misclassified.  It is clear that the performance is quite good, even
close to the plate limit at about 22nd.
}
\label{figGroth}
\end{figure}

\begin{figure}
\epsscale{0.5}
\plotone{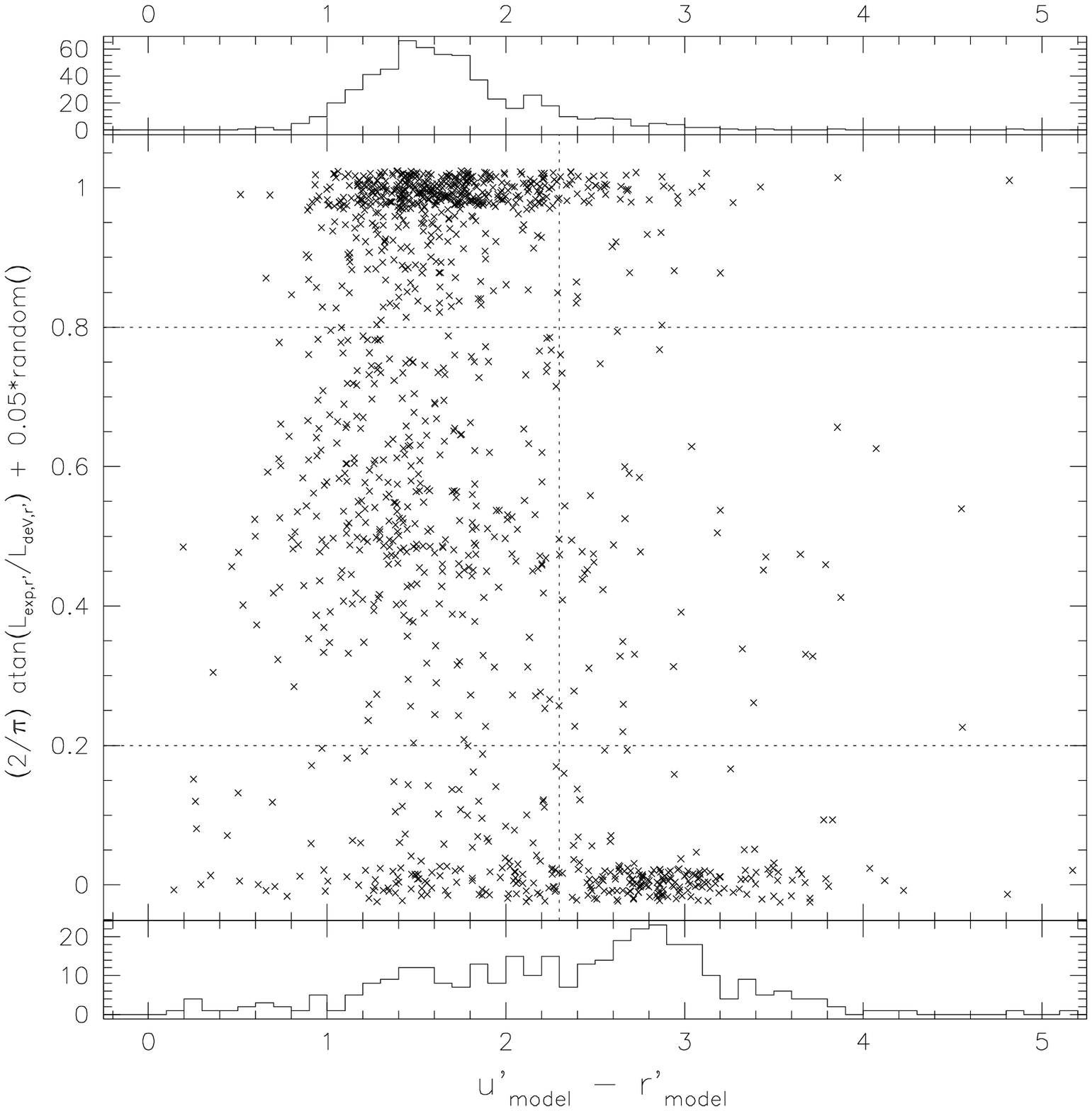}
\caption{
The relationship between morphological classification, based on the
ratio of the deVaucouleurs and exponential likelihoods.  The x-axis
is the $u'-r'$ colour, which divides the galaxies nicely into two
classes, presumably early- and late-type (Strateva et al. 2001).
The y-axis shows the
likelihood ratio (mapped into the range $[0,1]$); above and below the
plot are shown the marginal distributions of galaxies which lie
\emph{outside} the pair of dotted lines.  The correlation
of colour with morhology is clearly seen. Data is from a few square degrees of
run 745.}
\label{figGalClass}
\end{figure}

We fit three models to every object, in every band: a PSF, a pure
deVaucouleurs profile, and an exponential disk; the galaxy models are
convolved with the local PSF (as estimated using the KL expansion of
the previous section). This is potentially an expensive operation as
it involves a 3-dimensional ($r_e, a/b, \phi$) non-linear
minimisation; each iteration requires the calculation of a 2-d
analytical model of a galaxy followed by convolution with the PSF and
the calculation of $\chi^2$ by summing over many pixels of the image.
We make heavy use of pre-calculated tables of models, and pre-extract
the radial profile into a series of annuli, each containing 12
30$^\circ$ sectors; in consequence, fitting a single galaxy model in a
single band takes of order 1.5ms on an 800MHz alpha.

The primary use of these models is in star/galaxy separation and
morphological classification of galaxies.  We initially hoped to use
the relative likelihoods of the PSF and galaxy fits to separate stars
from galaxies, but found that the stellar likelihoods were tiny for
bright stars, where the photon noise in the profiles is small, due to
the influence of slight errors in modelling the PSF.  Instead we found
the ratio of the \textit{flux} in the best-fit galaxy model to
that in the PSF to be an excellent discriminant.

Figure \ref{figGR} shows a colour-magnitude diagram from a small area
of SDSS imaging data.
The top left panel shows only objects classified as stars; note that
most objects with colours of $g' - r' \approx 0.9$ are preferentially
classified as galaxies.  The star/galaxy separation is independent
of the object's colours, so this rejection \emph{must} be a measure
of how well the star/galaxy classification is working.

Studies of the performance of the SDSS S/G separation in
the Groth strip data (where accurate classification is available from
HST imaging) indicate that separation is reliable to at least a $r'$
of 21.5 in data that has a $5\sigma$ limit of $r' \approx 22$.

The $u' - r'$ colour of galaxies is a good discriminant of Hubble type
(Strateva et al. 2001).  Figure \ref{figGalClass} shows  $u' - r'$
plotted against what is essentially the likelihood ratio for
deVaucouleurs and exponential models shows that the galaxy likelihoods
provide clear \textit{morphological} classification to $r' \approx 20$,
in data with a PSF $5\sigma$ limit of about 22.5.

\section{Conclusions and Software Sociology}

As far as he knows, this section represents the views only of the
primary author and not of his coauthors.  Those of you who know
him will have heard these opinions before.

The SDSS has been very challenging technically, scientifically, and
managerially.  In all categories the software stands out: The hardest
technical aspect of building the SDSS was probably the software,
although building the mosaic camera wasn't easy; some of the software
was a major \textit{scientific} challenge; and the software was
undoubtedly the hardest part of the project to manage.

Let me expand upon some of these issues.  We have found it extremely
hard to hire good people to work on astronomical software.  There is
no career path within the universities for software specialists,
despite the fact that there's no logical distinction between building
hard- and soft-ware instruments. Smart and sensible graduate students,
desirous of a career in astronomy, simply don't choose to specialise
in the software required to reduce modern observational datasets.

Hiring computer professionals is not the solution to this problem.
Besides being (if competent) too expensive for the average
astronomical project, they simply don't possess the skills needed to
solve the \emph{scientific} challenges posed by astronomical data.  We
need \emph{scientists} to resolve scientific problems, albeit with
support from people whose job it is to know about optimizers, LALR(1)
grammars, and good software engineering practices.  We also need our
software-scientists to be in rich scientific environments, where they
can talk with (e.g.) the quasar-scientists about the data analysis that
they are carrying out.

If we, as a community, knew how to reuse software from one project on
another some of these problems might be alleviated, but I don't
believe that they would go away.  The availability of good numerical
libraries hasn't made the development of new cosmological codes stop;
the impetus for change comes from the desire to do things better, not
just from the not-invented-here syndrome.

I believe that part of the problem is that we, as a community have not
yet faced the reality that software is \emph{difficult}, and that the
dynamic range between the really good and the average programmer is
as great as that between Lyman Spitzer and the average graduate
student.  This makes management difficult; imagine trying to get a
collaboration of 100 self-opinionated astronomers to agree about the
best way to solve a problem, and tell me why this is any easier than
running a large modern collaboration involving large amounts of
software.  I reluctantly believe that we must learn to run large
software projects (and all large projects nowadays are large software
projects) as benevolent dictatorships --- of course with the implicit
hope that I shall be the dictator (but not the manager).

\acknowledgments

The Sloan Digital Sky Survey (SDSS) is a joint project of The
University of Chicago, Fermilab, the Institute for Advanced Study, the
Japan Participation Group, The Johns Hopkins University, the
Max-Planck-Institute for Astronomy, New Mexico State University,
Princeton University, the United States Naval Observatory, and the
University of Washington. Apache Point Observatory, site of the SDSS
telescopes, is operated by the Astrophysical Research Consortium
(ARC).

Funding for the project has been provided by the Alfred P. Sloan
Foundation, the SDSS member institutions, the National Aeronautics and
Space Administration, the National Science Foundation, the
U.S. Department of Energy, Monbusho, and the Max Planck Society.

The SDSS Web site is http://www.sdss.org/.

%
%


\end{document}